\documentclass[preprint2,longabstract]{aastex}

\begin{document}

\title{Orbital Parameters for the Black Hole Binary XTE J1650-500$^1$}

\author{Jerome A. Orosz}
\affil{Department of Astronomy, San Diego State University,
5500 Campanile Drive, San Diego, CA 82182-1221}
\email{orosz@sciences.sdsu.edu}

\author{Jeffrey E. McClintock}
\affil{Harvard-Smithsonian Center for Astrophysics, 60 Garden Street,
Cambridge, MA 02138}
\email{jem@cfa.harvard.edu}

\author{Ronald A. Remillard}
\affil{Center for Space Research, Massachusetts Institute of Technology,
Cambridge, MA 02139-4307}
\email{rr@space.mit.edu}

\author{St\'ephane Corbel}
\affil{Service d'Astrophysique, 
DAPNIA, b\^{a}iment 709, l'Orme des merisiers, 
CEA Saclay, 91191 Gif-sur-Yvette Cedex, France, and
Universit\'e Paris VII, 2 place 
Jussieu, 75251 Paris Cedex 05, France}
\email{corbel@discovery.saclay.cea.fr}

\altaffiltext{1}{Based on observations made with
the Magellan 6.5m Clay
telescope at Las Campanas Observatory of the Carnegie Institution and
on observations made with
European Southern Observatory (ESO) Telescopes at the Paranal Observatory 
under programme ID 
69.D-0644(A)}

\begin{abstract}
We present $R$-band photometry of the X-ray transient and candidate
black hole binary XTE J1650-500 obtained between 2003 May  and August
with the 6.5m
Clay Telescope.  A timing analysis  of these data reveals a photometric
period of $0.3205\pm
0.0007$ days (i.e.\ 7.63 hr) with a possible
alias at 0.3785 days (9.12 hr).  Our photometry
completely rules out the previously published spectroscopic
period of 0.212 days (5.09 hr).  Consequently, 
we reanalyzed the 15 archival ESO/VLT 
spectra
(obtained 2002 June by  
Sanchez-Fernandez et al.) that were the basis
of the previously published spectroscopic period.  We
used a 
``restframe search''
technique that is well suited for cases when the signal-to-noise
ratio of individual spectra is low.
For each of roughly 1.1
million binary ephemerides, we summed all of the spectra in a trial
restframe of the secondary star, and each restframe spectrum was
cross-correlated against a template spectrum. 
We then searched for the set of orbital parameters that
produced the strongest cross-correlation value. The results confirmed
the photometric period of 0.3205 days, and rule out the alias
period near 0.38 days.
The best value for the velocity semiamplitude of the companion
star is $K_2 = 435 \pm 30$
km s$^{-1}$, and
the corresponding optical mass function is $f(M)=2.73 \pm 0.56\,M_{\odot}$.
The spectral type of the companion star is not well constrained
because we only have six template spectra available to us.  The K4V
template provides the best match; next best matches are provided by
the G5V and K2III templates.
We also 
find that the accretion disk dominates the light in the $R$-band
where the disk fraction is 80\% or higher, although this value
should be treated with caution owing to the poor signal-noise-ratio
and the limited number of templates.
The amplitude of the phased $R$-band light curve is 0.2 magnitudes,
which gives a lower limit to the inclination of $50\pm 3^{\circ}$
in the limiting case of no contribution to the $R$-band light curve
from the accretion disk.
If the mass
ratio of XTE J1650-500 is similar to the mass ratios of other
black hole binaries like A0620-00 or GRS 1124-683 (e.g.\
$Q\gtrsim 10$), then  our lower limit to the inclination  gives
an upper limit to 
the mass of the black hole in 
XTE J1650-500  of $M_1 \lesssim 7.3\,M_{\odot}$.  
However, the mass can be considerably
lower if the $R$-band flux is dominated by the accretion disk.  For
example, if the accretion disk does contribute 80\% of the flux, as our
preliminary results suggest, then the black hole mass would be only
about $4\,M_{\odot}$.

\end{abstract}

\keywords{binaries: spectroscopic --- black hole physics ---
stars: individual (XTE J1650-500) --- X-rays: stars}

\section{Introduction}

X-ray novae (XN) provide the strongest evidence for the existence
of stellar mass black holes.  XN are interacting binaries where
a neutron star or a black hole accretes matter from a companion
(usually a dwarf-like K or M star).  
The accretion rate on to the black hole increases
substantially during the outburst phase, and hence the X-ray
luminosity can vary by large amounts  (e.g.\ 
factors of $10^6$ or
more).  The majority of XN spend most of their time in a ``quiescent''
state where the X-ray luminosity is on the order of the optical luminosity
of the companion star.  It is in quiescence  where the companion star
can be best studied in the optical, where the observed radial
velocity and light curves of the companion lead to dynamical
mass measurements for the compact primary.  
There are 18 X-ray binaries (15 of them XN) where
the
mass of the primary has been shown to exceed
the maximum mass of a stable neutron star
($\approx 3\,M_{\odot}$, Kalogera \& Baym 1996),
confirming the presence of black holes in these
systems: 
GRO J0422+32 (Orosz \& Bailyn 1995, Filippenko, Matheson, \& Ho 1995);
A0620-00 (McClintock \& Remillard 1986); 
GRS 1009-45 (Filippenko et al.\ 1999);
XTE J1118+480 (McClintock et al.\ 2001, Wagner et al.\ 2001);   
GS 1124-683 (Remillard, McClintock, \& Bailyn 1992);
4U 1543-47 (Orosz et al.\ 1998); 
XTE J1550-564 (Orosz et al.\ 2002);
GRO J1655-40 (Bailyn et al.\ 1995);
GX 339-4 (Hynes et al.\ 2003);
H1705-250  (Remillard et al.\ 1996);
SAX J1819.3-2525 (Orosz et al.\ 2001);
XTE J1859+226 (Filippenko \& Chornock 2001);
GRS 1915+105 (Greiner, Cuby, \& McCaughrean 2001);
GS 2000+25 (Casares, Charles, \& Marsh 1995);
GS 2023+338 (Casares, Charles, \& Naylor 1992);
Cyg X-1 (Gies \& Bolton 1986); 
LMC X-3 (Cowley et al.\ 1983); 
LMC X-1 (Hutchings et al.\ 1987).
These sources
open up the possibility of studying general relativity
in the strong field regime.  For example, the study of
high frequency quasiperiodic oscillations in the X-ray light curves
of certain XN may lead to a measurement of black hole spin
(e.g.\ McClintock \& Remillard 2004 and cited references).  We must
press hard to obtain further observations of black hole masses in order
to fully pursue these opportunities.

XTE J1650-500 (hereafter J1650) was discovered
by RXTE on 2001 September 5 (Remillard 2001) and subsequently reached
a peak X-ray intensity of 0.5 Crab.
Based on subsequent
observations, J1650 was established as a strong black
hole candidate based on its X-ray spectrum and variability
in the X-ray light curve (Markwardt, Swank, \& Smith 2001; Revnivtsev
\& Sunyaev 2001; Wijnands, Miller, \& Lewin 2001).  
The radio counterpart was discovered with the Australia Telescope
Compact Array (ATCA) by Groot et al.\ (2001). Further radio observations
sampled the behavior of XTE J1650-500 along all its X-ray
states (Corbel et al.\ in preparation).
We highlight
two key results obtained during its outburst phase.
First, RXTE observations during the third and fourth weeks of the
outburst yielded a strong X-ray QPO with an rms amplitude of
$5.0 \pm 0.4\%$ at a frequency of $\nu=250\pm 5$ Hz 
(Homan et al.\ 2003b).
Second, {\em XMM-Newton} observed a broad, skewed emission
line due to Fe K$\alpha$ (Miller et al.\ 2002).
Those authors argue that their results
imply the primary is a nearly maximal Kerr black hole that is
delivering its spin-down energy to the accretion flow.

The first significant observational program in the optical
was that of
Sanchez-Fernandez et al.\ (2002, hereafter SF2002).  
They observed J1650 on the
night of 2002 June 10 with the fourth 8.2m telescope at
the European Southern Observatory, Paranal,
and reported the following orbital elements:
an orbital period of $P=0.212\pm 0.001$ days and a velocity semiamplitude
of $K_2=309\pm 4$ km s$^{-1}$, resulting in an optical mass
function of $f(M)=0.64\pm 0.03\,M_{\odot}$.
The results in this paper contradict these findings.

In this paper we report the results of our photometric study of
J1650.  A time series analysis of our photometry rules out the
orbital period  reported by SF2002.  
Consequently we also report
herein our reanalysis of the SF2002 data obtained from the ESO archives.
We show that the spectroscopic period we derive from these data
is consistent with our photometric period  $P=0.3205$ days,
and we go on to determine the
orbital elements of the system.  We outline below our
observations and reductions, and our analysis techniques.  We end
with a brief discussion of the implications of our results regarding
the fast QPO  observed for J1650.

\section{Observations}

\subsection{Optical Photometry}

\begin{figure*}
\epsscale{1.80}
\plotone{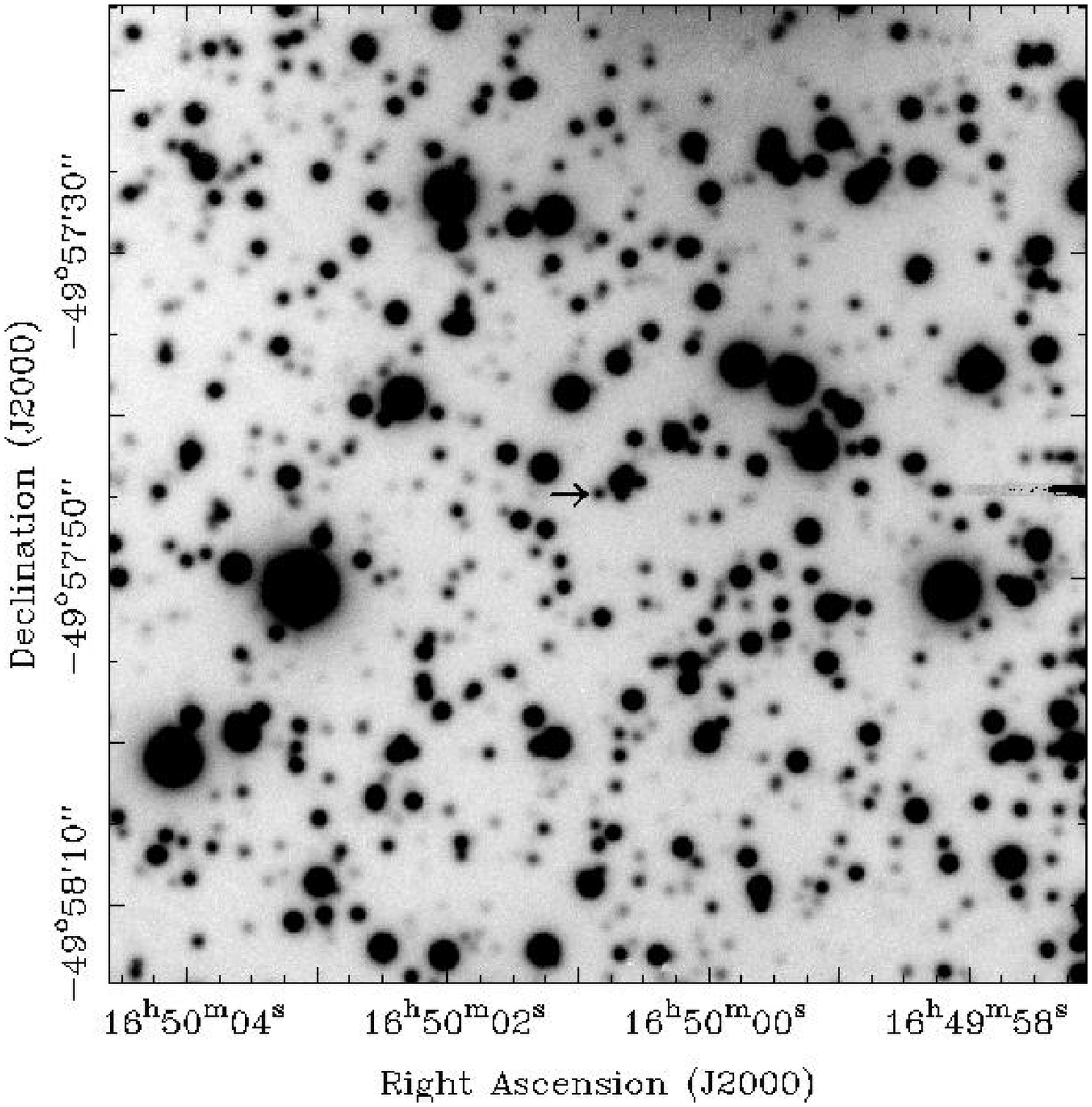}
\caption{An $R$-band finding chart for J1650.  The field of view
is one arcminute by one arcminute.
}
\label{figfc}
\end{figure*}

We observed J1650 with the 6.5m Clay telescope at Las Campanas
Observatory of the Carnegie Institution 2003 May 31 and June 1
using the Magellan Instant Camera (MagIC) and an $R$-band filter.
A total of 47 $R$-band images with typical exposure times between
500 and 600 seconds
were obtained over the two nights in good conditions: the average
seeing was $\approx 0.6$ arcsec and it was photometric about 75\% of the
time.
M. Holman and P. Schechter kindly provided an additional
65 images obtained 2003 August 1 and 2 using the same instrumentation,
and the observing conditions were comparable to those
just described.
In order to minimize readout time, the MagIC camera is read out
simultaneously using four amplifiers.  Thus, bias and flat-fielding is
done separately for each quadrant of the detector.  These calibrations
and the merging of the quadrant images into a single image were
performed using the publicly-available MagIC reduction pipeline.
Stetson's programs DAOPHOT IIe, ALLSTAR, and DAOMASTER
\citep{ste87,ste90,ste91,ste92a,ste92b}
were used to extract the stellar intensities and derive
the light curve for J1650
(see Orosz \& Wade 1999
for a detailed discussion of the overall
procedure used).  
This
suite of codes gives robust instrumental magnitudes.  DAOPHOT fits the
point spread function (PSF)
for each image using several relatively isolated
bright stars, and ALLSTAR uses the PSF and finds instrumental
magnitudes for all of the
stars on an image simultaneously (local background subtraction
is included).  Finally DAOMASTER iteratively solves for zero-point
shifts in the magnitude scales from image to image by essentially
using all of the stable stars as ``comparison'' stars.  Thus
the final
results are insensitive to changes in the seeing and to changes in the
sky transparency.
Fig.\ \ref{figfc} shows a finding chart for J1650 made from
the MagIC data.

\subsection{Optical Spectroscopy}

SF2002 observed J1650 on the
night of 2002 June 10 with the fourth 8.2m telescope at
the European Southern Observatory, Paranal,
using the FORS2
imaging spectrograph with an MIT CCD mosaic, the 1200R grism,
the GG435 blocking filter, and a fixed slit width of 0.7 arcsecond.
This instrumental configuration yields a wavelength range
of 5750-7310 \AA\ with a resolution of about 0.76 \AA\ FWHM.
SF2002 obtained a total of 15 spectra of J1650 in somewhat
marginal seeing (typically between 1.0 and 1.5 arcseconds).
They also obtained the spectra of six bright G- and K-type dwarfs
using the same instrumental configuration.
We obtained these data from the ESO archive, and used tasks in 
IRAF\footnote{IRAF is distributed by the National Optical Astronomy
Observatories,
which are operated by the Association of Universities for Research
in Astronomy, Inc., under cooperative agreement with the National
Science Foundation.} to
perform the CCD reductions and to extract the spectra.  Since the
flat field and wavelength calibration images were obtained exclusively
during the
daytime hours at Paranal, it was necessary to 
correct for flexure in the spectrograph by 
applying small corrections ($\le 0.6$\AA)
to the wavelength scales of the extracted spectra
using bright night-sky emission lines.  

\section{Analysis}

\begin{figure*}
\epsscale{1.80}
\plotone{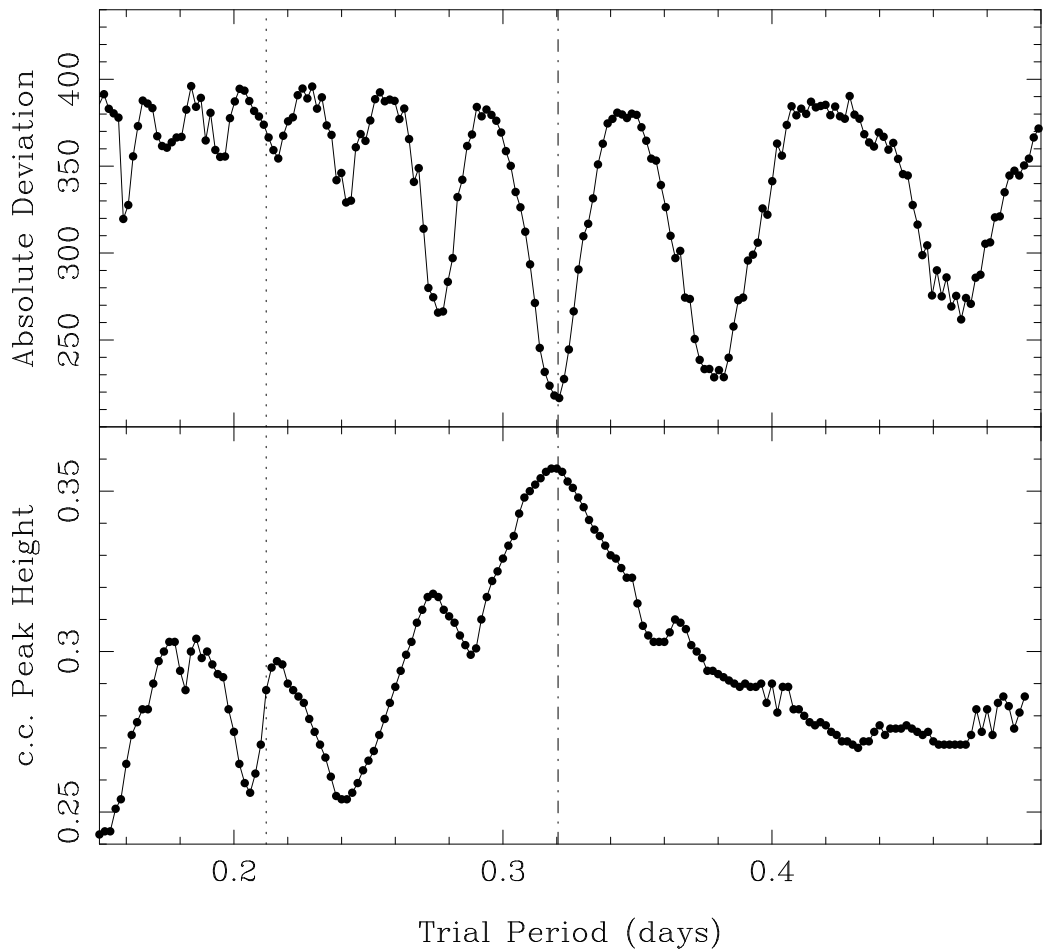}
\caption{The results from our search for the binary period are
presented.  On the top, the periodogram made from the J1650 photometry
using ELC is shown.  This plot shows the absolute deviation as a
function of the trial period, where a smaller value of the absolute
deviation indicate a better fit.  On the bottom, the periodogram for
J1650 derived from the ``restframe'' analysis of the spectra is shown.
This plot shows the maximum cross-correlation value as a function of
the trial period, where a higher cross-correlation value indicates a
better fit.  Both periodograms have their best statistic at the same
trial period, namely $P=0.320$ days (dot-dashed line).  The dotted
line denotes the period reported by SF2002 ($P=0.212$ days).}
\label{fig1}
\end{figure*}

The Magellan light curve of J1650 showed considerable variability, so
we phased the data on the period reported by SF2002 ($P=0.212$ days).
To our surprise, the light curve phased on that period showed
considerable scatter.  Consequently, we searched the Magellan
light curve for periodicities 
using
the ELC code with its genetic fitting algorithm (Orosz \&
Hauschildt 2000; Orosz et al.\ 2002).  About 1,106,000
ellipsoidal models were generated and compared with the data using
the absolute deviation as the merit function:
$$
A=\sum_{i=1}^N\left\vert {y(x_i;a)-y_i\over \sigma_i} \right\vert,
$$
where  $y(x_i;a)$ denotes the model value computed at
$x_i$, $y_i$ is the observed quantity at the same $x_i$, and
$\sigma_i$ is the uncertainty in $y_i$.  
This merit function is more robust than the $\chi^2$ function because it
is less sensitive to a few outlying points.
Models were computed using a period range
of 0.15 to 0.50 days.  After all of the fits were performed,
the merit function hypersurface was projected onto
the trial period axis
(see Orosz et al.\ 2002 for a more in-depth discussion of this error
estimation technique);
the resulting periodogram is shown in the top of 
Fig.\ \ref{fig1}.  The absolute deviation has a minimum value at a
trial period of 0.3205 days, and a secondary minimum at a trial period
of 0.3785 days.  

\begin{figure*}
\epsscale{1.80}
\plotone{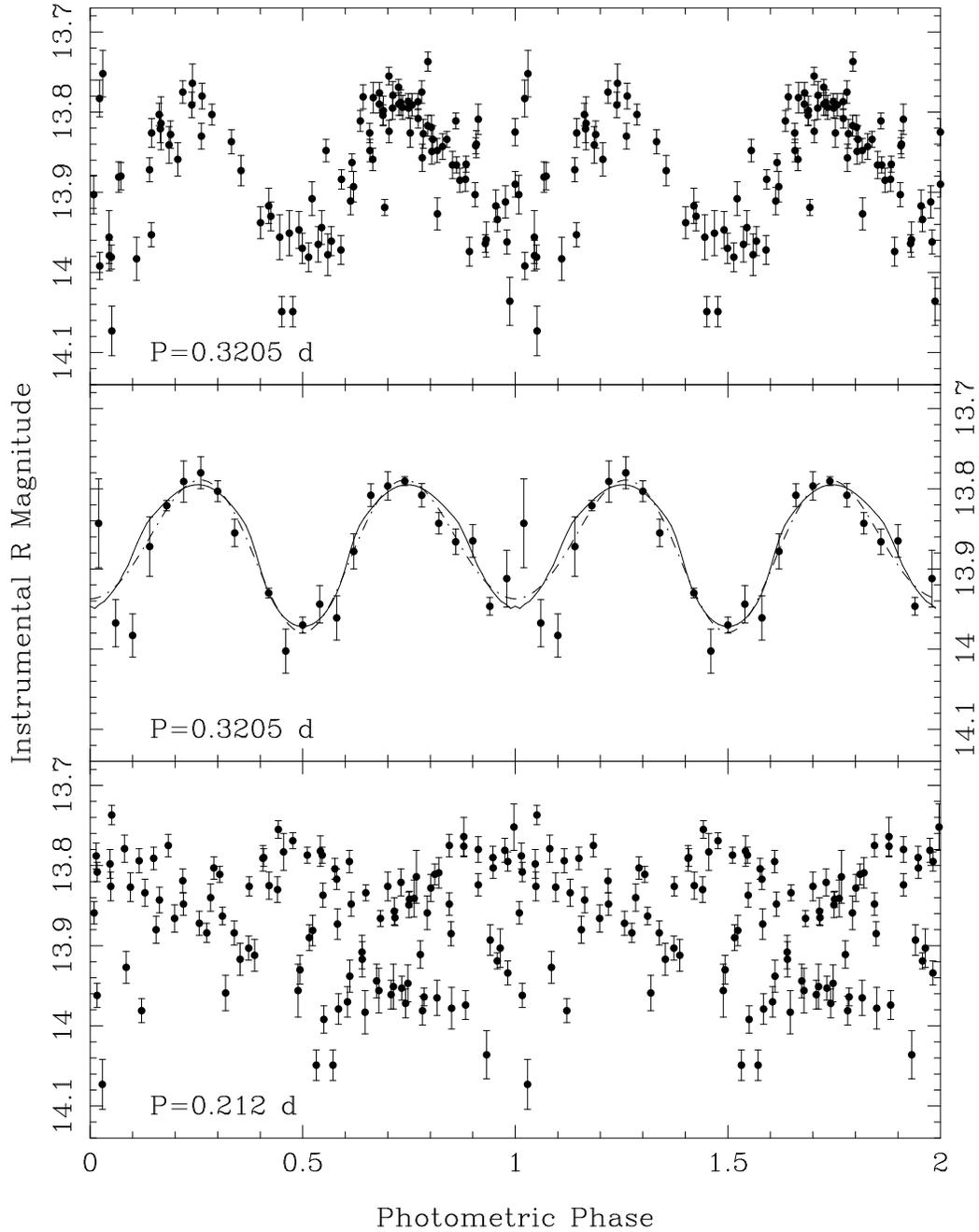}
\caption{Top: The Magellan light curve of J1650 phased on the
photometric period of $P=0.3205$ days.  Middle: The phased light
curved smoothed using 20 bins in phase, along with two representative
ellipsoidal model fits (see Table \ref{tabparm} for model parameters)
The solid line shows a model with an accretion that dominates the
light (denoted at model A in Table \ref{tabparm}) and the dashed line
shows a model with no accretion disk (model B in Table \ref{tabparm}).
Bottom: The Magellan light curved folded using the Sanchez-Fernandez
period of $P=0.212$ days.
}
\label{fig2}
\end{figure*}

Fig.\ \ref{fig2} (top)
shows the Magellan light curve phased
on a period of $P=0.3205$ days.
There is no dip in the absolute deviation near the
period reported by SF2002 ($P=0.212$ days), and the light
curve phased on that period is essentially a scatter plot
(bottom of Fig.\ \ref{fig2}).
As a check on our results,
we used the `pdm' task in IRAF, which is
an implementation
of the phase dispersion technique of Stellingwerf (1978).   A similar
periodogram was obtained.
To estimate the error on our photometric measurement of the orbital period,
we did additional fits using ELC's genetic algorithm and
the normal $\chi^2$ as the merit function.  We find a $1\sigma$ uncertainty
of about 0.0007 day, so we adopt $P=0.3205\pm 0.0007$ day.

We then analyzed the VLT spectra in an attempt to understand the
disagreement of our photometric period with the spectroscopic period
reported by SF2002.  We tried to extract radial velocities
using the standard cross-correlation technique
of Tonry \& Davis (1979, implemented in the IRAF `fxcor' task).  
However, we
found that the spectra were very noisy and yielded only a few marginal
measurements of velocity. 
We therefore used the  ``restframe'' analysis that
we developed for H1705-250 (Remillard et al.\ 1996).
This technique is similar to the ``skew mapping'' technique sometimes
used for cataclysmic variables (Smith, Cameron, \& Tucknott 1993;
Vande Putte et al.\ 2003), and is quite simple to employ.  Suppose
one has a time series of
spectroscopic observations.  If the spectra are Doppler shifted to zero
velocity and coadded using the  correct orbital elements
(e.g.\
the period $P$, the
semiamplitude $K_2$ and the time of maximum velocity $T_0$), 
then the
absorption features of the companion star will appear at the same
wavelengths in all of the individual spectra.  Consequently, the lines 
in the
summed spectrum will have a higher signal-to-noise ratio than in
the individual spectra.
If, on the other hand, one uses the incorrect orbital elements,
then the absorption lines from the companion will be at different
wavelengths in different individual spectra, and hence will
be ``averaged out'' in the
resulting summed spectrum.  

In our
implementation of the restframe analysis, 
we used a
FORTRAN program that writes IRAF scripts to do the Doppler shifting and
the spectrum summation.
The fxcor task was used to provide a measure of how well a
restframe spectrum matched the template spectrum.  In particular,
we used the peak cross-correlation value as the measure of the
goodness-of-fit of the template spectrum to the restframe
spectrum.  The wavelength region used for the cross-correlation
analysis was 5907-6261\AA, 6304-6513\AA, 
6610-6800\AA, and 7000-7239\AA.
This region excludes telluric lines, and the H$\alpha$
emission line.

We are interested in three orbital elements: the period $P$, the
semiamplitude $K_2$ and the time of maximum velocity $T_0$.
This three dimensional parameter space was searched by making
two dimensional grids of restframe
spectra in the $K_2-T_0$ plane at several different values
of the period $P$.  Specifically,
we 
used
periods in the range 0.1500 to 0.5000 days in steps of
0.0020 days.  The restframe spectra for each trial period were stored in a
separate subdirectory on disk.  For each period, we constructed a
grid of restframe spectra in the
$K_{2}-T_{0}$ plane as follows. The range considered for $K_{2}$ was
300--600 km s$^{-1}$ in steps of 2 km s$^{-1}$.  For $T_{0}$ the range was
centered on HJD 2,452,436.6 and the step size was 0.005 days.  The
extent of the range for $T_{0}$ was adapted to accommodate the trial
value of the period.  An individual restframe spectrum was generated
for each set of orbital elements.    In total we
generated 1,166,146 restframe spectra using this same number of
scripts stored as separate files.  

Next, we did a cross-correlation analysis on the restframe
spectra.  It was therefore necessary to find a good template
spectrum.
To do this, we adopted
a period of 0.320 days found from the photometry
and considered the restframe spectra
in the $K_2-T_0$ plane at this period. 
We then cross-correlated these restframe spectra against all six of
the template spectra in turn and determined that the K4V star BS5568
provided the best match since it gave the strongest
cross-correlation peaks.
Next,
all of the restframe spectra at all periods were
cross-correlated in batch mode against the K4V template.
We made a periodogram
by parsing 
the fxcor log files to get the peak 
cross-correlation value for
each restframe spectrum.  The maximum peak 
cross-correlation value within each period subdirectory
was saved, resulting in the periodogram shown at the bottom of
Fig.\ \ref{fig1}.  The maximum cross-correlation value occurs
for a trial period of $P=0.320$ days, in agreement with the
photometric results, which thereby establishes the orbital period.  
Thus, we rule out the alias 
photometric period
near 0.38 days
and again rule out the 0.212 day spectroscopic period reported by SF2002
(Fig.\ \ref{fig1}).  We also made a periodogram using a template
with spectral type G5V.  Overall, the cross-correlation values were
much lower and the periodogram was much noisier.  However, there still
was a peak near $P=0.320$ days, and we conclude our adopted
spectroscopic period is not sensitive to the choice of the template.

\begin{figure*}
\epsscale{1.80}
\plotone{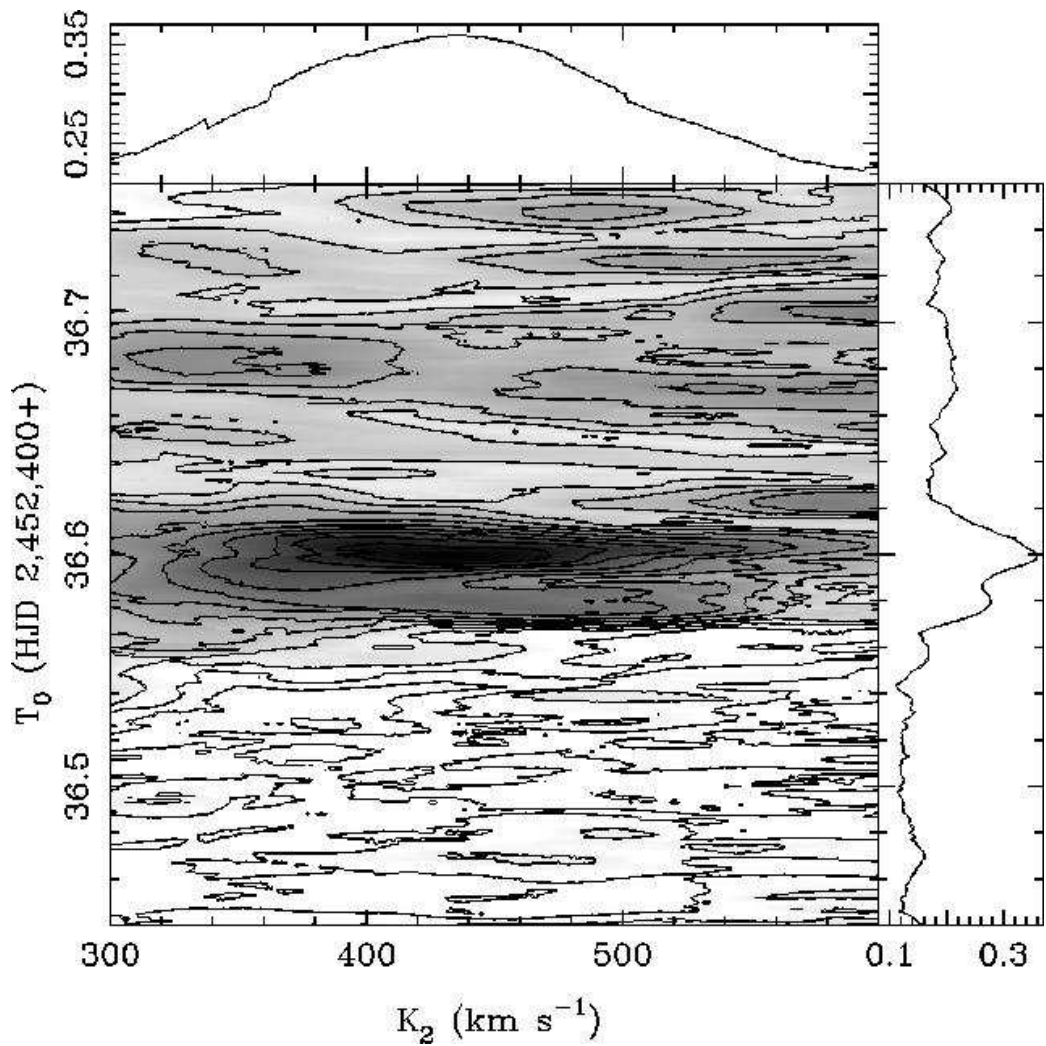}
\caption{A grayscale image and contour plot showing the peak
cross-correlation value as a function of $K_2$ and $T_0$ at a fixed
period of $P=0.32$ days.  The maximum cross-correlation values happen
when $430\lesssim K_2\lesssim 439$ km s$^{-1}$ and $T_0=2,452,436.600$
(HJD).  The thinner panels show the cross cuts of the
cross-correlation values through the peak along the $K_2$-axis and the
$T_0$-axis.}
\label{fig3}
\end{figure*}

Adopting a period of $P=0.320$ days, we performed a finer
search of the $K_2-T_0$ plane, with step sizes of 1 km s$^{-1}$ in
$K_2$
and 0.001 days in $T_0$.  The results are shown in Fig.\ \ref{fig3}.
The cross-correlation values have a fairly well-defined peak 
value
along the $T_0$
axis, where the maximum cross-correlation is for $T_0=
2,452,436.600$ (HJD).  On the other hand, the peak along the $K_2$ axis
is broad with a maximum that occurs in the range
$430 \lesssim K_2 \lesssim 439$ km s$^{-1}$.  
Judging from the width of
the peak and on our experience with H1705-250 (Remillard et al.\
1996), we adopt $K_2=435 \pm 30$ km s$^{-1}$.  The optical mass function is
then
$$
f(M)={PK_2^3\over 2\pi G} = 2.73\pm 0.56\,M_{\odot}.
$$

\begin{figure*}
\epsscale{1.80}
\plotone{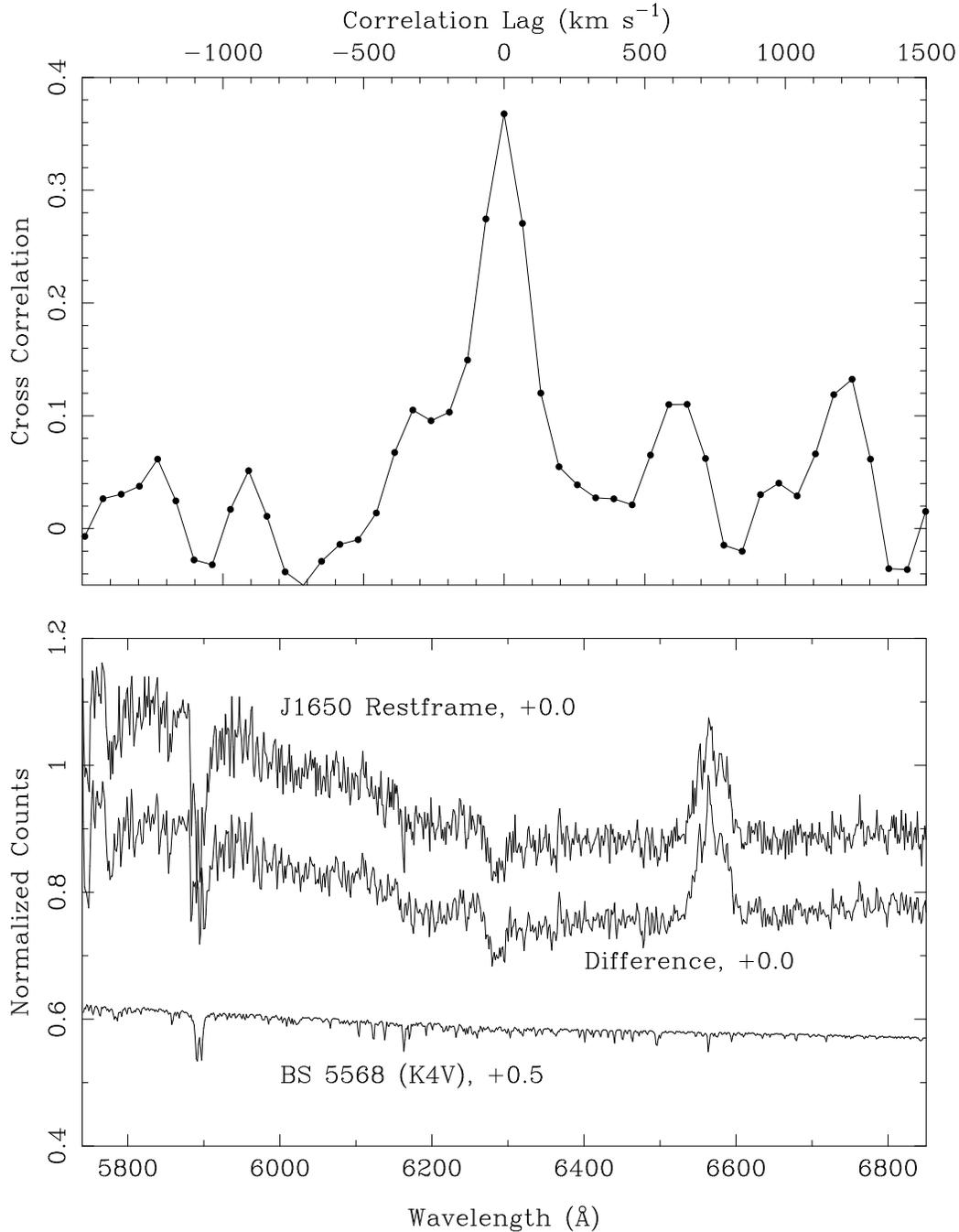}
\caption{Bottom:  The best normalized restframe spectrum of J1650
(top spectrum), the normalized spectrum of the K4V star BS 5568,
scaled by 0.1 and shifted upwards by 0.5 (bottom spectrum), and
the difference spectrum (middle spectrum).
Top:  The cross-correlation function of
the best restframe spectrum against the K4V star BS 5568.
}
\label{plotspect}
\end{figure*}

The best restframe spectrum found for the finer grid is shown in
Fig.\ \ref{plotspect}.  Although few obvious K-star absorption features
are apparent, the cross-correlation of this restframe spectrum using
the K4V star BS 5568 as a template does yield a significant peak near
zero velocity (Fig.\ \ref{plotspect}).   

Since the spectroscopic period agrees with the photometric period,
our initial determination that the K4V template provides the
best match needs no refinement.  In order to be more quantitative, 
we cross-correlated the best restframe spectrum from the fine grid
against all six template spectra.  Table \ref{tabcor} 
gives the cross-correlation
value at the peak and the Tonry \& Davis `$r$' value (which is
a measure of the significance of the peak) for each template.
As noted earlier, the K4V template has the best cross-correlation, so
we adopt a spectral type of K4V for the companion star.
However, one should note that the GV5 star
BS 7330 provides the second-best match, and as such our assignment of
a spectral type of K4V should be treated with caution. 
We attempted to decompose
the restframe into its stellar and accretion disk components using
the technique outlined in Marsh, Robinson, \& Wood (1994).
Using the K4V template BS 5568, we found  that the accretion disk
dominates in the $R$-band, where its contribution is $\gtrsim 80\%$.
However, owing to the limited number of templates and the poor
signal-to-noise, this result should also be treated with caution.

Our measurement of the optical mass function sets a lower limit on the
mass of the compact primary.
To find the actual mass we need to find the 
mass ratio of the two components
and inclination of the orbital
plane.
If the companion star fills its Roche lobe and is
in synchronous rotation, the mass ratio can be computed if
one can measure $K_2$ and the projected rotational velocity of
the companion star (e.g.\ Wade \& Horne 1988).  One
can also get an estimate of the mass ratio if the radial
velocities of the H$\alpha$ emission line can be reliably
measured (e.g.\ Orosz et al.\ 1994).  However, in the case of J1650,
the poor quality of the spectroscopic
data prevents us from making either measurement.
We can use our Magellan light curve
to place some constraints on the inclination.  However, the results
are sensitive to the amount of light the accretion disk contributes in the
$R$-band, and our estimate of the disk contamination outlined above
is quite uncertain.  
Nevertheless, if we assume the limiting
case of no light from the disk in the $R$-band, then we find an inclination
of $i=50\pm 3^{\circ}$ (see Fig.\ \ref{fig2} for a representative
model for this case).  
If we add a substantial contribution from
the disk in the $R$-band ($\gtrsim 80\%$), then the inclination is
$70^{\circ}$ or higher (a representative ellipsoidal model for this case
is also shown in Fig.\ \ref{fig2}).     
Owing to the lack of X-ray eclipses, the
inclination is less than about $80^{\circ}$ (the exact limit
depends on the mass ratio).   If, for the sake of discussion, we assume
the
mass ratio of J1650 is similar to that of A0620-00 or GRS 1124-683
($Q\approx 10$), then an inclination of $50^{\circ}$ combined
with our measured value of the optical mass function yields
a black hole mass of $7.3\pm 0.6\,M_{\odot}$.  Likewise,
an assumed inclination of $70^{\circ}$ yields a black hole
mass of $4.0\pm 0.6\,M_{\odot}$.

\section{Discussion}

Is general relativity (GR) the correct theory of gravity in
the strong fields found near a black hole?  One promising
approach to answering this fundamental question is offered
by the key discovery of the {\em Rossi X-ray Timing Explorer}
(RXTE) that seven Galactic black holes (including J1650)
display quasiperiodic
(QPO)
X-ray oscillations in the range of $100-450$ Hz
(McClintock \& Remillard 2004 and cited references).
These fast QPOs must be produced near the event horizon
since the X-rays originate there and since such
high frequencies are comparable to the Kepler frequency of the
innermost stable circular orbit around a 
black hole ($\nu_K=2199-16150\,(M/M_{\odot})^{-1}\,{\rm Hz}$ depending
on the dimensionless spin parameter $a_*$,
where the full range of $a_*$
is 0 to 1
e.g.\
Kato, Fukue, \& Mineshige 1998).

RXTE has made a further important discovery:
four of these seven black holes produce pairs of stationary,
fast QPOs that have frequencies in a 3:2 ratio (McClintock
\& Remillard 2004; Homan et al.\ 2003a).
Such commensurate frequencies are a hallmark
of non-linear resonance phenomena.  Thus,
this discovery has promoted a ``resonance
model'' that invokes enhanced emissivity at the radius in the accretion
disk where two of the three spatial coordinate frequencies 
(e.g.\ Keplerian and radial) are commensurate
(Abramowicz \& Kluzniak 2001;
Abramowicz et al.\ 2003).  For the three black holes 
with the QPO pairs {\em and} measured masses
(e.g.\ GRO J1655-40, XTE J1550-564, and GRS 1915+105),
the QPO frequency is correlated with the black
hole mass, where $\nu\propto M^{-1}$ (e.g.\
McClintock \& Remillard 2004).
This scaling is expected for GR oscillations, but different sources
can lie on the same curve only if they have similar values of the
dimensionless spin parameter $a_*$.

Schnittman \& Bertschinger (2004) performed
ray tracing calculations in the Kerr metric 
for emitting blobs orbiting a black hole at a radius where
there is a 3:1 resonance between
the azimuthal ($\Omega$)
and radial coordinate frequencies.  
It was shown that general relativistic effects may impart a 
QPO at
$\Omega$ and a beat oscillation at $\slantfrac{2}{3}\, \Omega$ with
a relative strength that depends on the angular width of the emitting
blob of material.  This interpretation, when combined with accurate
mass measurements, yields values for the dimensionless spin
parameter:
$0.40 \le a_*\le 0.55$ for GRO J1655-40 and $0.3 \le a_* \le 0.6$
for XTE J1550-564 (Remillard et al.\ 2002).  This illustrates the potential
diagnostic power of fast QPOs, if we can specify the correct oscillation 
mechanism.  Furthermore, measurements of black hole mass
and QPO frequency allow {\em model-independent}
comparisons of spin differences between black hole subclasses
as distinguished, for example, by relativistic jets or binary
period.  

With our measurement of the orbital period and optical mass
function of J1650, we have taken the first steps need to fully exploit
the potential of this system.  If the QPO at $\nu=250\pm 5$
Hz observed by Homan et al.\ (2003b) represents the $2\nu_0$
oscillation, then the predicted mass of the black hole would
be about $7.5\,M_{\odot}$, if J1650 lies on the same $\nu:M$ curve
on which GRO J1655-40, XTE J1550-564, and GRS 1915+105 lie
(i.e.\ if the black holes all have similar values of the
dimensionless spin parameter $a_*$).
Although there are substantial
uncertainties in both our measured value for the optical mass
function and in our inclination estimate, it appears that the
black hole mass might be less than $7.5\,M_{\odot}$, which
in the context of the orbital resonance model would imply that the
value $a_*$ in J1650 is much
smaller than it is for the other three sources
(e.g.\ Abramowicz et al.\ 2004).  
On the other hand, on the basis
of X-ray spectroscopy, Miller et al.\ (2002) have argued that J1650
contains a maximal Kerr black hole.  Thus J1650 may be a challenging case
for the orbital resonance model of the high frequency QPOs,  
although we again point out that our mass estimate
has large uncertainties.

Recent models for black hole formation (e.g.\
Fryer \& Kalogera 2001) predict a continuous and
roughly exponential distribution
of masses for black holes in binary systems, where
the number of black holes formed falls off as the mass
increases.  Depending on their model assumptions, these
authors predict the number of black holes with masses
in the range  $3-5\,M_{\odot}$ should be roughly two to three
times the number of black holes with masses in the range 
$8-10\,M_{\odot}$.  Bailyn et al.\
(1998) performed a statistical analysis of seven black holes with
dynamical mass estimates and found strong evidence for a peak
in the mass distribution centered near $7\,M_{\odot}$ and
a ``gap'' in the distribution between the neutron star
masses and the peak near $7\,M_{\odot}$.  
Since that time, the sample of objects has more than doubled, and it
now appears that the evidence for a peak in the distribution near 
$7
\, M_{\odot}$ is weak (see the references cited in \S1 and Orosz 2003).
However, black holes with masses in the $3-5\,M_{\odot}$
range still seem to be rare, with
GRO J0422+32 being
the only  good candidate 
(Gelino \& Harrison 2003). In this regard, J1650 is an interesting
system because of its probable low mass.  

Clearly additional data will be required to make more definitive statements
about the mass of the compact object in J1650.
It should not be unduly difficult to obtain higher quality spectra. 
In hindsight, we now know that the companion star is a  K-star,
and as such it should have relatively strong absorption features
near 5170\AA.  This region was not included in the VLT spectra
owing to the relatively high resolving power of the 1200R grism.
It should be easier to measure radial velocities
in spectra with more wavelength coverage (at the expense of resolving
power).  Also, spectral observations obtained in good to excellent
seeing conditions ($\lesssim 0.6$ arcseconds)
obviously will have better signal-to-noise than
the current spectra do (the seeing varied between 1 and about 1.4
arcseconds).  Additional photometry should be obtained, and it
would be helpful if two or more bandpasses could be used, since the
amount of contaminating light from the disk should vary with color.  

\section{Summary}

Using $R$-band photometry of XTE J1650-500 obtained with the
6.5m Clay telescope, we have measured a photometric
period of $P=0.3205\pm 0.0007$ days.  This value 
is confirmed by our reanalysis of archival spectroscopic data.  
Our reanalysis of the spectroscopy also yields a velocity 
semiamplitude
for the companion star of $K_2=435\pm 30$ km s$^{-1}$, which when
combined with the orbital period, gives an optical mass function
of $f(M)=2.73\pm 0.56\,M_{\odot}$.  
The spectral type of the companion star is not well constrained
because we only have six template spectra available to us.  A
template spectrum with a spectral type of K4V
provides the best match; next best matches are provided by
template spectra with spectral types of G5V and K2III, respectively.
The summed spectrum suggests that the accretion disk
may contribute a substantial
amount of light in $R$ ($\approx 80\%$ or more).
Using the $R$-band light curve,
we find a lower limit to the inclination of $i=50\pm 3^{\circ}$,
which gives an upper limit to the mass of the black hole
of $M_1\le 7.3\,M_{\odot}$,
assuming a mass ratio of $Q=10$.  
If we assume that the accretion disk contributes 80\%
of the light in $R$, then the inclination would be about $70^{\circ}$
or more, and the black hole mass would be about $4\,M_{\odot}$ (again
assuming a mass ratio of $Q=10$).

\acknowledgments

We thank Matt Holman and Paul Schechter for providing supplementary
photometric data and Mauricio Navarette for help with the
observations.  This work was supported in part by NASA grant 
NAG5-9930.


\clearpage

\clearpage

\clearpage

\clearpage

\clearpage

\clearpage

\begin{deluxetable}{rcc}
\tablecaption{Ellipsoidal Model Parameters\label{tabparm}}
\tablewidth{0pt}
\tablehead{
\colhead{Parameter\tablenotemark{a}} &
\colhead{Model A} &
\colhead{Model B} 
}
\startdata
$i$ (deg)    &  79.0  &   51.7 \cr
$Q$          &  20.7  &   1.1     \cr
$T_1$ (K)    &  4500  & 4500     \cr
$\beta$      &  0.099 & 0.099    \cr
$T_{\rm disk}$ (K)  &  15000  & \nodata \cr
$r_{\rm inner}$  &  0.01   & \nodata \cr
$r_{\rm outer}$  & 0.99    &  \nodata \cr
$\xi$            &  -0.3   & \nodata  \cr   
$\beta_{\rm rim}$ (deg)            &  2.0   & \nodata  \cr   
\enddata

\tablenotetext{a}{See Orosz \& Hauschildt (2000) for a
full discussion of the parameters and their use in the
ELC code.}

\end{deluxetable}

\begin{deluxetable}{rrcc}
\tablecaption{Cross-Correlation Values for Templates\label{tabcor}}
\tablewidth{0pt}
\tablehead{
\colhead{Star} &
\colhead{Spectral type} &
\colhead{Cross-Correlation} &
\colhead{Tonry \& Davis $r$}  \\
\colhead{} &
\colhead{} &
\colhead{Peak Value} &
\colhead{}
}
\startdata
BS 5568   &  K4V     & 0.36  & 7.69  \cr
BS 7330  &  G5V      & 0.27  & 6.84  \cr
HD 11301  &  K2III   & 0.26  & 3.91  \cr
BS 2668   & K0V      & 0.18  & 1.88  \cr
BS 8042  &  G3 IV    & 0.10  & 1.22  \cr
HD 209290  &  M0V    & 0.06  & 0.70  \cr
\enddata


\end{deluxetable}

\end{document}